\documentclass{PoS}

\newcommand{\gl}{\tilde{g}}
\newcommand{\sq}{\tilde{q}}
\newcommand{\sqb}{\bar{\tilde{q}}}
\newcommand{\as}{\alpha_{\mathrm{s}}}
\newcommand{\si}{\sigma}
\newcommand{\sih}{\hat \sigma}
\newcommand{\shat}{{\hat s}}

\newcommand{\rhohat}{{\hat \rho}}
\newcommand{\onebf}{{\bf 1}}
\newcommand{\eigbf}{{\bf 8}}
\newcommand{\eigbfa}{{\bf 8_{\mathrm \bf A}}}
\newcommand{\eigbfs}{{\bf 8_{\mathrm \bf S}}}
\newcommand{\tptbf}{{\bf 10}\oplus {\bf \overline{10}}}
\newcommand{\twsbf}{{\bf 27}}
\newcommand{\ppbar}{\hspace{0.1em} p \hspace{-0.69em} ^{^{(-)}}}
\newcommand{\ppbart}{\hspace{0.1em} p \hspace{-0.55em} ^{^{\tiny (-)}}}
\def\nn{\nonumber}

\title{Soft gluon resummation for squark and gluino pair-production at hadron colliders}

\ShortTitle{Soft gluon resummation for squark and gluino pair-production at hadron colliders}

\author{Wim Beenakker\\
  Theoretical High Energy Physics, Radboud University Nijmegen, P.O. Box 9010\\
  NL-6500 GL Nijmegen, The Netherlands\\
        E-mail: \email{W.Beenakker@science.ru.nl}}

\author{Silja Brensing, Michael Kr\"amer, \speaker{Anna Kulesza}\\
        Institute for Theoretical Physics, RWTH Aachen University,
        D-52056 Aachen, Germany\\
        E-mail: \email{brensing@physik.rwth-aachen.de},
        \email{mkraemer@physik.rwth-aachen.de}, \email{anna.kulesza@physik.rwth-aachen.de}}

\author{Eric Laenen\\
        ITFA, University of Amsterdam, Valckenierstraat 65, 1018 XE Amsterdam, The
  Netherlands \\
  ITF, Utrecht University, Leuvenlaan 4, 3584 CE Utrecht, The Netherlands\\
  Nikhef Theory Group, Science Park 105, 1098 XG Amsterdam, The Netherlands\\
        E-mail: \email{t45@nikhef.nl}}

\author{Leszek Motyka\\
        II Institute for  Theoretical Physics, University of Hamburg, Luruper Chaussee 149, D-22761, Germany \\
Institute of Physics, Jagellonian University, Reymonta 4, 30-059
Krak\'{o}w, Poland\\
        E-mail: \email{motyka@mail.desy.de}}

\author{Irene Niessen\\
  Theoretical High Energy Physics, Radboud University Nijmegen, P.O. Box 9010\\
  NL-6500 GL Nijmegen, The Netherlands\\
        E-mail: \email{i.niessen@science.ru.nl}}

\abstract{We report on the study of soft gluon  effects in 
the production of squarks and gluinos at hadron colliders. Close to
production threshold, the emission of soft gluon results 
  in the appearence of large logarithmic corrections in the theoretical
  expressions. In order to resum these corrections at
  next-to-leading-logarithmic accuracy appropriate one-loop anomalous
  dimensions have to be calculated. We present the calculation of the
  anomalous dimensions for all production channels of squarks and
  gluinos and provide numerical predictions for the Tevatron and the LHC. }

\FullConference{RADCOR 2009 - 9th International Symposium on Radiative Corrections (Applications of Quantum Field Theory to Phenomenology) \\
		 October 25-30 2009\\
		 Ascona, Switzerland}

\begin{document}

\section{Introduction}
Extensions of the Standard Model (SM) involving supersymmetry (SUSY) are one of the
best motivated models of new physics. 
The searches for signals of supersymmetry 
are currently carried out by the Tevatron experiments~\cite{CDF,D0} and in the coming
years will be undertaken by the experiments at the Large Hadron
Collider (LHC). Within the Minimal Supersymmetric 
Standard Model (MSSM)~\cite{mssm}, the dominant production processes
of supersymmetric particles (sparticles) at hadron colliders
are those involving pairs of coloured sparticles, i.e. squarks  
and gluinos, in the final state. 
Depending on the outcome of the experimental 
searches, predictions for the total rates 
for these production processes are either used
to draw the exclusion limits for the mass parameters or will
help to determine the masses of the sparticles.
It is therefore important to know the cross sections for
hadroproduction of squarks and gluinos with high theoretical accuracy.

There are four pair-production processes of squarks and gluinos in
hadronic collisions: $p  \ppbar \to \sq \sqb \,, p \ppbar \to \sq \sq\,, p \ppbar
\to \gl \gl\,,  p \ppbar \to \sq \gl  \,.
$
The next-to-leading order (NLO) SUSY-QCD corrections~\cite{BHSZ1,BHSZ2} to
hadroproduction processes are generally positive at the central
scale. Depending
on the process and the masses of sparticles considered, they can be large. 
As pointed out in~\cite{BHSZ2}, an important
part of the contributions to the hadronic cross sections comes from
the energy region near the partonic production threshold, reached when the partonic 
center-of-mass (c.o.m.) energy squared $\shat$ approaches $4\,m^2$, 
where $m$ is the average mass of the produced particles. 
In this region, two types of corrections dominate: the
Coulomb corrections, due to exchange of gluons between
slowly moving massive particles, and the soft-gluon corrections, due to
emission of low-energetic gluons off the coloured initial and
final states.  
The large size of the soft-gluon emission contributions can be traced
back, for the perturbative $n$-th order correction,  to the logarithmic terms of the form $\as^n \log^{k}(\beta^2)$ where
$k=2n, \dots, 0$ with $\beta \equiv \sqrt{1 - 4 m^2/\shat}$.
The effects of
the soft-gluon emission are taken into account to all orders in
perturbation theory by performing 
resummation of the threshold logarithms.

Here we review calculations of the resummed total cross sections
for the squark and gluino hadroproduction presented in~\cite{KM1,KM2,BBKKLN}. The LO Coulomb corrections
to $\sq\sqb$ and $\gl\gl$ production have been resummed
in~\cite{KM2}. For the squark-antisquark
production process the dominant contribution to the
next-to-next-to-leading order (NNLO) correction coming from the
resummed cross section at next-to-next-to-leading-logarithmic (NNLL)
level has been studied in~\cite{Langenfeld:2009eg}.
Recently, a general formalism allowing for simultaneous resummation of threshold and
Coulomb corrections has been presented and applied to $\sq\sqb$ production~\cite{BFS}.

\section{NLL threshold resummation}

The inclusive hadroproduction cross section $\si_{h_1h_2\rightarrow
  kl}$ for two massive SUSY particles $k$ and $l$, where $k,l$ can be
a squark ($\tilde{q}$), antisquark ($\bar{\tilde{q}}$) or gluino
($\tilde{g}$), can be written in terms of its partonic version
$\si_{ij\rightarrow kl}$ as
\begin{eqnarray}
  \si_{h_1 h_2 \to k l}\bigl(\rho, \{\underline{m}^2\}\bigr) 
  \;=\; \sum_{i,j} \int d x_1 d x_2\,d\hat{\rho}\;
        \delta\left(\hat{\rho} - \frac{\rho}{x_1 x_2}\right)
        f_{i/h_{1}}(x_1,\mu^2 )\,f_{j/h_{2}}(x_2,\mu^2 )\,
        \si_{ij \to kl}\bigl(\hat{\rho},\{ \underline{m}^2\},\mu^2\bigr)\,,\nn \\
\label{eq:hadronicxs}
\end{eqnarray}
where $\{\underline{m}^2\}$ denotes all masses entering the calculations, $i,j$
are the initial parton flavours, $f_{i/h_1}$ and $f_{j/h_2}$ the
parton distribution functions, and $\mu$ is the common factorization
and renormalization scale. For the hadronic
c.o.m. energy $\sqrt S$, we define the threshold variable
$\rho\;=\;(m_k+m_l)^2/S\;=\; m^2/S$. It 
measures the distance from threshold  for inclusive
production of two final-state particles with masses $m_k$ and $m_l$ in terms of the energy fraction.
The partonic equivalent of this threshold variable is defined as
$\hat{\rho}=\rho/(x_1x_2)$, where $x_{1,2}$ are the momentum fractions
of the partons.
The resummation of threshold logarithms is performed in the
space of Mellin moments $N$ of the cross section~(\ref{eq:hadronicxs}), taken with respect to
the variable $\rho$. Since the soft radiation carries colour charge, it can change the
colour state of the underlying hard scattering. Consequently, the
components in the resummed formula describing 
hard-scattering as well as the components 
corresponding to soft emission carry dependence on the colour configuration
of all four particles participating in the reaction $ij \to kl$. This
dependence can be vastly simplified if one performs the calculation in an
orthogonal basis in the colour space for which the soft anomalous
dimension matrices, governing the soft emission, are diagonal. In this case, 
up to next-to-leading-logarithmic (NLL)
accuracy, the resummed partonic cross section reads~\cite{Sdy,CTdy,Contopanagos:1996nh,
  Kidonakis:1997gm,Kidonakis:1998bk,Bonciani:1998vc}, 
\begin{eqnarray}
&&\si^{\rm (res)}_{i j \to kl}(N, \{\underline{m}^2\}, \mu^2) = \int_0^1 d \rhohat \; \rhohat^{N-1}\;
           \si_{i j\to kl}\bigl(\rhohat,\{ \underline{m}^2\}, \mu^2\bigr)= \nn \\
&& = \sum_{I}  
\tilde{\sih}^{(0)}_{ij\to kl,I}(N,\as(\mu^2))
\Delta_i (N+1,Q^2,\mu^2)  \Delta_j (N+1,Q^2,\mu^2) \Delta^{\rm
  (int)}_{ij\to kl,I} (N+1,Q^2,\mu^2).
\label{eq:nspace}
\end{eqnarray}

The functions $\Delta_i
(N+1,Q^2,\mu^2)$ resumming the collinear or collinear and soft radiation from the incoming
partons are universal and their form can be found
e.g. in~\cite{KM2}. The new elements
which have to be calculated in order to obtain the resummed predictions from~(\ref{eq:nspace})
are the LO partonic cross-sections in $N$-space,  $\tilde{\sih}^{(0)}_{ij\to kl,I}(N,\as(\mu^2))$, 
and the soft function $\Delta^{\rm
  (int)}_{ij\to kl,I} (N+1,Q^2,\mu^2)$.  Both types of contributions
are dependent on the colour configuration of the underlying
hard-scattering $ij \to kl$ and
have to be calculated separately for each colour exchange channel,
indicated by the index $I$. We perform the calculations in the $s$-channel orthogonal colour
basis.

The soft function, describing large-angle soft
gluon emission, at the NLL accuracy is given by
$$
\ln \Delta^{\rm (int)}_{ij\to kl,I,N} = 
 \int_0^1  dz \frac{z^{N-1}-1}{1-z} \frac{\alpha_s(4m^2(1-z)^2)}{\pi} D_{ij\to
kl,I} . 
$$
The coefficients $D_{ ij\to kl,I}$ are related to the threshold limit
$\beta \to 0 $ of the one-loop anomalous dimension matrices
$\Gamma_{ij \to kl}$. In this limit, the matrices calculated in the $s$-channel orthogonal colour
basis become
diagonal. Each diagonal element  $\Gamma_{ij \to kl, II}$
contributes to the  corresponding $D_{ ij\to kl,I}$ coefficient.
The expressions for $\Gamma_{ij \to kl}$ in the case
$kl=\sq\sqb$ can be found in~\cite{Kidonakis:1997gm}, $kl=\gl\gl$ in~\cite{KM1,KM2}
and $kl=\sq\sq, \sq\gl$ in~\cite{BBKKLN}. The corresponding
expressions for the LO $N$-space colour contributions and the $D_{ij
  \to kl,I}$ coefficients have been calculated in~\cite{KM1,KM2} and
  in~\cite{BBKKLN} for the $\sq\sqb$, $\gl\gl$ and $\sq\sq$,$\sq\gl$
  production, respectively. We list the $D_{
    ij\to kl,I}$ coefficients  for all the subprocesses $ij \to kl$ of
  interest in Table 1. The values of the coefficients
  equal the negative values of the quadratic Casimir operators
  belonging to the irreducible SU(3) representations for the outgoing
  state in a given process. This confirms the physical picture of the
  soft gluon radiation at production threshold being governed by the total
  colour charge of the heavy-particle pair in the final state~\cite{BFS,Bonciani:1998vc}.
\begin{table}
\begin{center}
\begin{tabular}{ll}
$D_{q\bar q \to \sq \sqb,\,I} ^{(1)} = \{0,-3\}$    &  $I=\{\onebf,\eigbf\}$  \\
$D_{gg \to \sq \sqb,\,I} ^{(1)} = \{0,-3\}$ &  $I=\{\onebf,\eigbf\}$ \\
$D_{q\bar q \to \gl\gl,\,I} ^{(1)} = \{0,-3,-3\}$ &
$I=\{\onebf,\eigbfs,\eigbfa \}$  \\
$D_{gg\to \gl\gl,\,I} ^{(1)} = \{0,-3,-3,-6,-8\}$ &  $I=
\{\onebf,\eigbfs,\eigbfa,\tptbf,\twsbf \}$  \\
$D_{qq\to \sq \sq,\,I} ^{(1)} = \{-4/3, -10/3\}$ &   $I=\{ \bf{\bar
  3}, \bf{6}\}$    \\
$D_{qg\to \sq \gl,\,I} ^{(1)} = \{-4/3, -10/3, -16/3 \}$ & $I=\{ \bf{3},\bf{\bar
 6},\bf{15} \}$
 \end{tabular}
\end{center}
\caption{The soft coefficients $D_{ij\to kl,I}$ together with the
  corresponding irreducible SU(3) representations spanned by the
  vectors of the colour basis.}
\end{table}

\section{Numerical results}

The resummation-improved cross sections, $\sigma_{\rm NLL+NLO}$, are obtained through matching
the NLL-resummed results with the complete NLO cross sections, in the
manner described e.g. in Ref~\cite{KM2,BBKKLN}. The NLO cross sections are
evaluated using the {\tt PROSPINO} code~\cite{prospino}, based on
calculations~\cite{BHSZ1,BHSZ2}. We assume the
left-handed and right-handed squarks in the final state to be mass-degenerate and sum
over chiralities. No top squarks in the final state are considered.
In our calculations we use the MSTW 2008 NLO pdfs~\cite{mstw}
and two-loop running strong coupling constant $\as$ in the $\overline{\rm MS}$  scheme
with five active flavours. 

\begin{figure}
\begin{center}
\hspace{-1.0cm}
\begin{tabular}{ll}
(a)\includegraphics[width=.4\columnwidth]{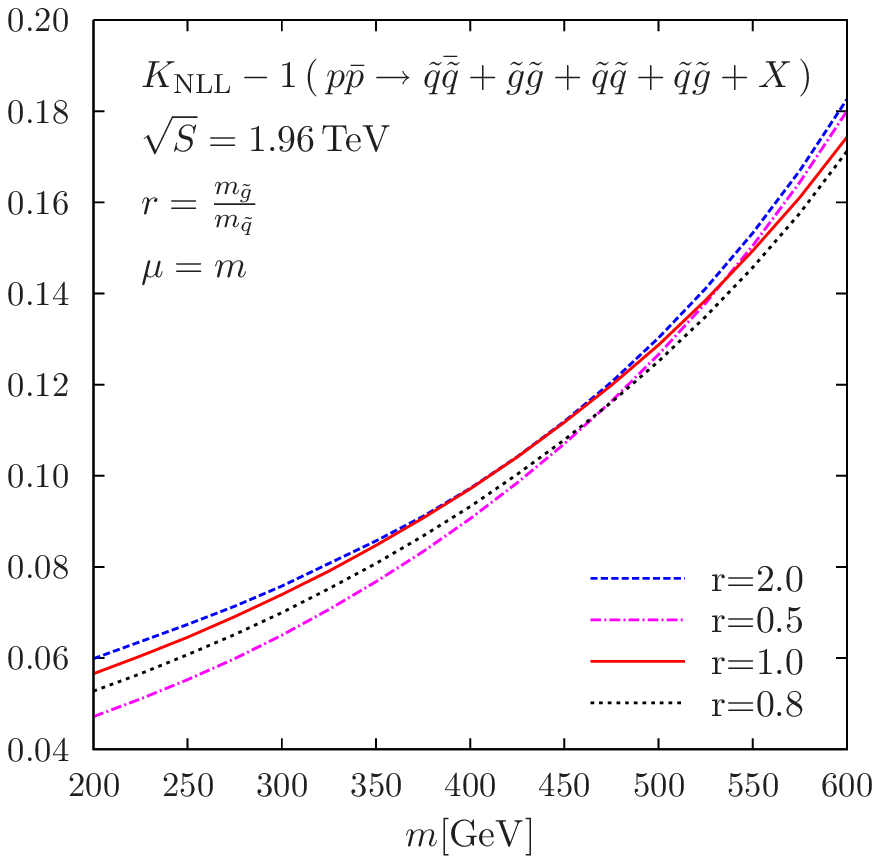} & 
(b)\includegraphics[width=.4\columnwidth]{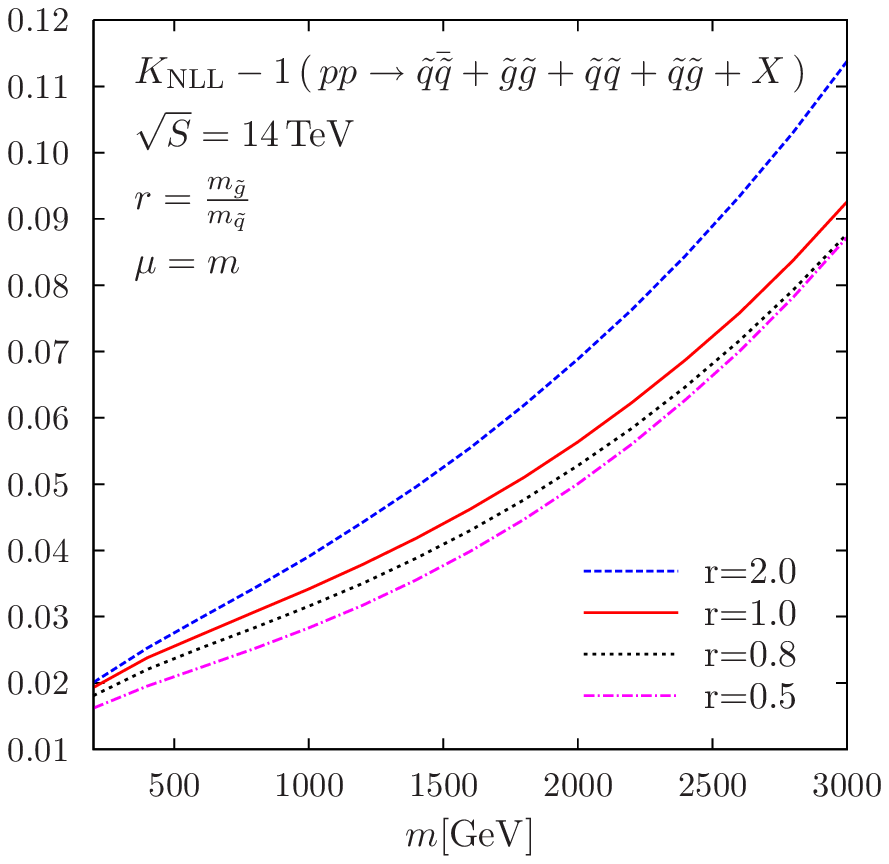}
\end{tabular}
\caption{The relative NLL $K$-factor $K_{\rm NLL} -1 =
  \sigma_{\rm NLL+NLO}/\sigma_{\rm NLO}-1$ for the inclusive squark
  and gluino pair-production cross section, $p\bar{p}/pp \to
  \tilde{q}\tilde{q}+
  \tilde{q}\bar{\tilde{q}}+\tilde{q}\tilde{g}+\tilde{g}\tilde{g} + X$,
  at the Tevatron (a) and the LHC (b) as a function of the average
  sparticle mass $m$. Shown are results for various mass ratios $r
  = m_{\tilde{g}}/m_{\tilde{q}}$.}
\label{fig:k_inclusive}
\end{center}
\end{figure}

We find that the NLL corrections to hadroproduction of squarks and gluinos are
positive and lead to higher predictions for the total cross sections
compared to the NLO results. In general, the most significant
effects arise for processes characterised by high contributions 
from gluon initial states and the presence of gluinos in the final
states. This is to be expected due to the high colour
charge and, correspondingly, the large values of the Casimir
invariants involved.
The impact of the NLL resummation on the total cross section
for the inclusive squark and gluino production,
$p \ppbart \to  \sq\sqb\,, \sq\sq\,, \sq\gl\,, \gl\gl
+ X$, is illustrated in Fig.~\ref{fig:k_inclusive}. The relative
K-factor $K_{\rm NLL}-1 \equiv \sigma_{\rm NLL+NLO}/\sigma_{\rm
  NLO}-1$, shown for the Tevatron in  Fig.~\ref{fig:k_inclusive}a) and
for the LHC  in  Fig.~\ref{fig:k_inclusive}b), is calculated for
various values of a fixed ratio $r$ of squark and gluino mass, $
r=m_{\tilde{g}} / m_{\tilde{q}}$. 
For the average mass $m = 600$~GeV the
correction to the inclusive cross section at the Tevatron due to NLL
resummation can be as high as 18\%. The inclusive corrections are
smaller at the LHC for sparticle masses below 3~TeV (see
Fig.~\ref{fig:k_inclusive}b). Given the sparticle mass ranges that we
consider, this is consistent with the fact that the distance from
threshold, i.e.~the value of the variable $1-\rho=1-4m^2/S$, is on
average larger at the LHC than at the Tevatron.

\begin{figure}
\begin{center}
\hspace{-1.0cm}
\begin{tabular}{ll}
(a)\includegraphics[width=.4\columnwidth]{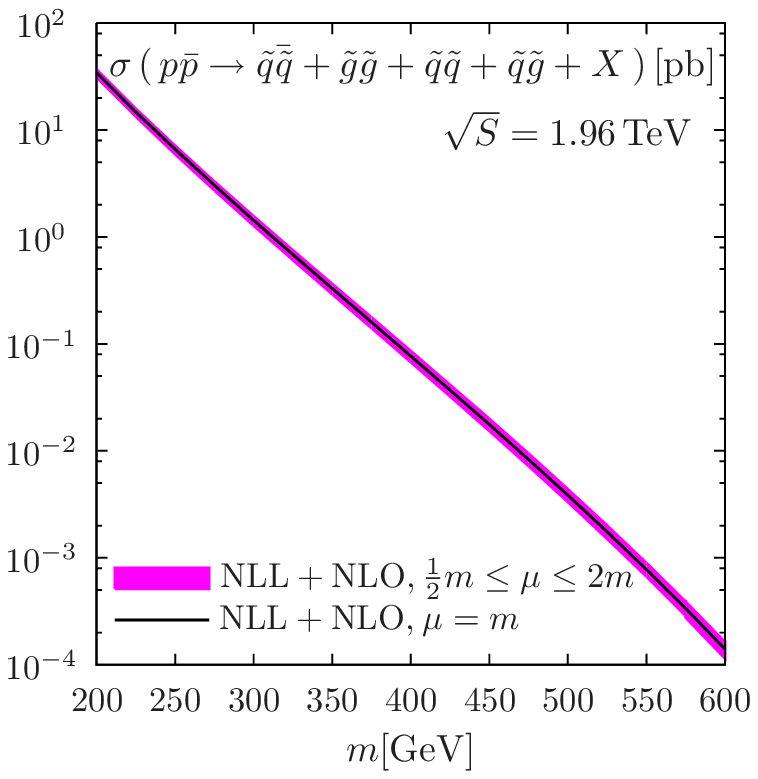}& 
(b)\includegraphics[width=.4\columnwidth]{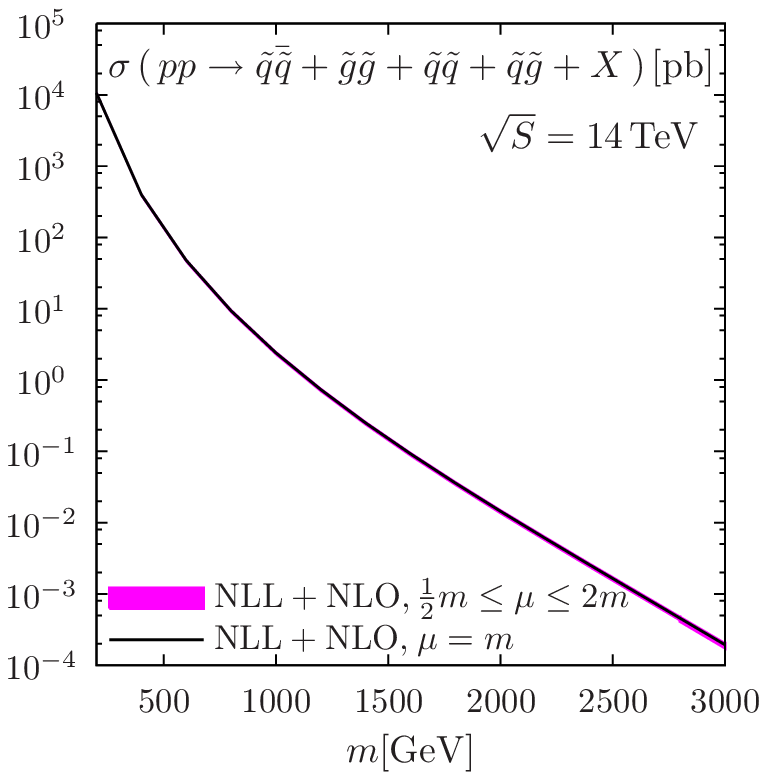}
\end{tabular}
\caption{The NLL+NLO cross section for inclusive squark and gluino
  pair-production, $p\bar{p}/pp \to \tilde{q}\tilde{q}+
  \tilde{q}\bar{\tilde{q}}+\tilde{q}\tilde{g}+\tilde{g}\tilde{g} + X$,
  at the Tevatron (a) and the LHC (b) as a function of the average
  sparticle mass $m$. Shown are results for the mass ratio $r =
  m_{\tilde{g}}/m_{\tilde{q}} =1$.  The error band corresponds to a
  variation of the common renormalization and factorization scale in
  the range $m/2\le\mu\le 2m$.}
\label{fig:total:matched}
\end{center}
\end{figure}

In Figs.~\ref{fig:total:matched}a) and \ref{fig:total:matched}b) we show
for the Tevatron and LHC, respectively, the resummed NLL+NLO total
cross section for inclusive squark and gluino production as a function
of the average sparticle mass $m$. For illustration we show these
results for the choice $m_{\sq} = m_{\gl}$. The error bands indicate
the theoretical uncertainty of the NLL+NLO total cross section due to
the scale variation in the range $m/2\le\mu\le 2m$. Both at the
Tevatron and at the LHC, soft-gluon resummation leads to a 
reduction by around 30\% in this part of the theoretical uncertainty. The results
presented in Fig.~\ref{fig:total:matched} are the most accurate
theoretical predictions currently available for the above processes.

\noindent
{\bf Acknowledgments}\\
This work has been supported in part by the Helmholtz Alliance
``Physics at the Terascale'', the DFG Graduiertenkolleg ``Elementary
Particle Physics at the TeV Scale'', the Foundation for Fundamental
Research of Matter (FOM), the National Organization for Scientific
Research (NWO), the DFG SFB/TR9 ``Computational Particle Physics'',
and the European Community's Marie-Curie Research Training Network
under contract MRTN-CT-2006-035505 ``Tools and Precision Calculations
for Physics Discoveries at Colliders''.

\end{document}